\documentclass[12pt]{article}
\usepackage{multirow}  
\usepackage{amsmath}
\usepackage{amssymb, latexsym}
\usepackage{amsthm}
\usepackage{caption}
\usepackage{setspace} 
\usepackage{xcolor}

\def\no{\noindent}
\allowdisplaybreaks

\usepackage{natbib} 
\setcitestyle{authoryear,round}

\usepackage{multirow, booktabs, threeparttable}
\newcommand{\pr}{{\rm pr}}
\newcommand{\e}{ { \mathbb{E}}}
\def\mS{\mathcal{S}}
\def\T{{ \mathrm{\scriptscriptstyle \top} }}

\newtheorem{theorem}{Theorem}[section]
\newtheorem{remark}[theorem]{Remark}
\newtheorem{proposition}[theorem]{Proposition}

\begin{document}

\centerline{\large {\bf Statistical Inference with Nonignorable Non-Probability Survey Samples}}

\bigskip 

\centerline{Yang Liu, \ Meng Yuan, \ Pengfei Li \ and \ Changbao Wu\footnote{Yang Liu is an Associate Professor at the School of Mathematical Sciences, Soochow University, Suzhou, Jiangsu 215006, China (Email: liuyang2023@suda.edu.cn). Meng Yuan contributed to this project as a postdoctoral researcher in the Department of Statistics and Actuarial Science at the University of Waterloo (Email: meng.yuan031@gmail.com). Pengfei Li and Changbao Wu are Professors in the same department at the University of Waterloo, Waterloo, Ontario N2L 3G1, Canada (Emails: pengfei.li@uwaterloo.ca and cbwu@uwaterloo.ca). This research was supported by the National Natural Science Foundation of China (Grant No. 12101239) and the Natural Sciences and Engineering Research Council of Canada.}}

\bigskip
\bigskip

\hrule

{\small
\begin{quotation}
\no
Statistical inference with non-probability survey samples is an emerging topic in survey sampling and official statistics and has gained increased attention from researchers and practitioners in the field. Much of the existing literature, however, assumes that the participation mechanism for non-probability samples is ignorable. In this paper, we develop a pseudo-likelihood approach to estimate participation probabilities for nonignorable non-probability samples when auxiliary information is available from an existing reference probability sample. We further construct three estimators for the finite population mean using regression-based prediction, inverse probability weighting (IPW), and augmented IPW estimators, and study their asymptotic properties. Variance estimation for the proposed methods is considered within the same framework. The efficiency of our proposed methods is demonstrated through simulation studies and a real data analysis using the ESPACOV survey on the effects of the COVID-19 pandemic in Spain.

\vspace{0.3cm}

\no
KEY WORDS \ Inverse probability weighting, nonignorable participation mechanism, outcome regression, pseudo likelihood, reference probability sample, variance estimation. 
\end{quotation}
}

\hrule

\bigskip
\bigskip

\section{Introduction\label{Section1}}

Survey sampling is an important branch of statistical science. It involves collecting and analyzing data from a finite population and has a wide range of applications in social, economical and health sciences and official statistics. Since the seminal work by \cite{neyman1934two}, probability sampling and design-based inference has been regarded as the gold standard in survey sampling \citep{kalton2019developments, wu2020sampling}. However, probability sampling has faced many challenges over the past three decades, including ever-increasing costs and non-response rates.

In the era of big data, convenient and inexpensive data are now available from non-probability samples such as web-panel surveys, administrative records, and other sources including social media content, web-scraped information, transaction records, sensor data, and satellite imagery. 
Non-probability samples can provide near real-time estimates, which contrasts sharply with the traditional approach of collecting data through probability samples \citep{beaumont2021pitfalls}. Statistical analysis with non-probability survey samples, however, presents distinct challenges compared to probability sampling methods \citep{baker2013summary}. While both probability and non-probability samples can suffer from biases due to undercoverage of certain population segments, the primary challenge with non-probability samples is the unknown participation mechanism. 

A popular framework for analysis of non-probability samples has been used in recent literature under two key assumptions: (i) the participation mechanism for the non-probability sample is ignorable; and (ii) auxiliary population information required for estimation can be obtained from an existing probability survey sample from the same population. 
Under this framework, several strategies have been developed to adjust for selection biases inherent in non-probability samples, thereby improving the validity of estimation and inferential procedures. One effective strategy is the model-based prediction approach, which assumes a shared parametric or nonparametric model for the outcome regression between the population and the non-probability sample. By using the estimated regression model from the non-probability sample, the population mean can be estimated using techniques such as mass imputation \citep{chen2020doubly, kim2021combining} or sample matching methods \citep{rivers2007sampling, yang2021integration}.
Another strategy involves the use of propensity scores, which correspond to the participation probabilities for the non-probability sample. A parametric model for the propensity scores is typically adopted and estimated using pseudo-likelihood methods \citep{valliant2011estimating, chen2020doubly, wang2021adjusted, chen2022pseudo}, as well as calibration weighting methods \citep{chen2020doubly, liu2023investigating}. Researchers have also explored non-parametric approaches using kernel and regression-tree-based techniques \citep{mercer2018selection, chu2019use}.
Using the estimated propensity scores, the inverse probability weighting (IPW) method can be applied to estimate the population mean \citep{chen2020doubly}.
To further improve estimation robustness, double robust inference has been developed by combining the aforementioned strategies \citep{chen2020doubly}. Bayesian methods have also been considered \citep{nandram2021bayesian, rafei2020big}. For comprehensive reviews, see \cite{yang2020statistical}, \cite{rao2021making}, \cite{kim2022gentle} and \cite{wu2022statistical}. 

The ignorability assumption used in the aforementioned framework, i.e., the participation probabilities for the non-probability sample do not depend on the response of interest given the observed covariates, may not hold in practice. For instance, we examine the effects of the COVID-19 pandemic on people's mood in Section \ref{Section4},   and evidences suggest that there is a positive correlation between good mood and participation as a positive mood fosters helping behaviour \citep{carlson1988positive, wolff2022day}.  Estimation results obtained under the ignorability assumption for such cases become unreliable and are typically biased. To the best of our knowledge, there is very limited research addressing the nonignorable participation mechanism for non-probability survey samples, with the exception of \cite{kim2023empirical}, who assumed that auxiliary variables are available for the entire finite population.

In this paper, we build upon the framework established by \cite{rivers2007sampling}, \cite{valliant2011estimating} and \cite{chen2020doubly}, where there is a non-probability sample with measurements on both responses and auxiliary variables, along with a reference probability sample that contains only auxiliary variables. We develop inferential procedures to estimate the population mean under a nonignorable participation mechanism. Our contributions include (1) the establishment of conditions for model identifiability under two assumptions similar to those in \cite{kim2023empirical}; 
%one involving logistic regression for the nonignorable participation mechanism, and the other assuming a parametric outcome model for the non-probability sample. 
%These conditions are met when an instrumental variable is included among the auxiliary variables. Second, 
(2) a proposed novel pseudo-likelihood method to estimate participation probabilities under the assumed nonignorable participation mechanism; and 
(3) the development of regression, IPW, and augmented IPW (AIPW) estimators of the population mean, and addressing the challenge of variance estimation for these estimators.

The rest of the paper is organized as follows. 
In Section \ref{Section2}, we presents our proposed inferential procedures for nonignorable non-probability survey samples. 
Section \ref{Section3} reports results from simulation studies to demonstrate the performance of our proposed methods. Section \ref{Section4} provides a real data analysis from the ESPACOV survey. Some additional remarks are given in Section \ref{Section5}.

\section{Inferential Procedures\label{Section2}}

In this section, we first introduce two parametric models for nonignorable participation and outcome regression for non-probability samples. We discuss issues with the identifiability of model parameters and develop a pseudo-likelihood method for parameter estimation. We further investigate the asymptotic properties and variance estimation of the regression, IPW, and AIPW estimators for the population mean.
\subsection{Problem setup} 

Let $\mathcal{U} = \{1, 2, \cdots, N\}$ denote a finite population comprising $N$ units. Attached to unit $i \in \mathcal{U}$ are values of auxiliary variables ${x}_i$ and a response variable $y_i$. Let $\mathcal{S}_A$ be a non-probability sample of size $n_A$ from $\mathcal{U}$ with an unknown participation mechanism and $\{(x_i,y_i),i\in \mathcal{S}_A\}$ be the sample dataset. 
Following \cite{rivers2007sampling}, \cite{valliant2011estimating} and \cite{chen2020doubly}, 
we assume the existence of a reference probability survey sample $\mathcal{S}_B$ with measurements $x_i$ but not $y_i$. The data structure can be represented as: 
\begin{equation}
    \left\{(x_i, y_i), i \in \mathcal{S}_A\right\} \cup \left\{(x_i, d_i^B), i \in \mathcal{S}_B\right\}\,,
    \label{np.data.structure}
\end{equation}
where $d_i^B$ are the known survey weights for the reference probability sample $\mathcal{S}_B$.

Let $R_i = I\left(i \in \mathcal{S}_A\right)$ denote the indicator variable representing whether unit $i$ is included in the sample $\mathcal{S}_A$. 
%We consider $\{(x_i, y_i, R_i), i \in \mathcal{U}\}$ as a random sample from a sub-population $(x, y, R)$. 
The participation probability $\pr(R=1 \mid x, y)$, also known as the propensity score, depends on both the auxiliary variables $x$ and the response variable $y$. Throughout this paper, we assume that participation probabilities are non-zero for all units. We impose two parametric models, as discussed in \cite{kim2023empirical}.
\begin{itemize}
\item[(i)]
The participation probability is described by a logistic regression model given as follows:
\begin{equation}\label{eq:model_prop_score}
\pi^A(x, y; \theta) =\pr(R=1 \mid x, y)= \frac{1}{1 + \exp (\alpha + x^\T\beta + \gamma y )},
\end{equation}
where $\theta = (\alpha, \beta^\T, \gamma)^\T$ is the vector of unknown model parameters. Cases with $\gamma = 0$ represent an ignorable participation mechanism, while $\gamma \neq 0$ indicates a nonignorable participation mechanism. 
 
\item[(ii)]
 The conditional probability density or mass function for the response $y$ given the auxiliary variables $x$ for units in the non-probability sample $\mathcal{S}_A$, denoted as $\pr(y \mid x, R=1)$, has a parametric form denoted as $f(y \mid x; \xi)$, where $\xi$ is a vector of unknown model parameters.
\end{itemize}
Our objective is to conduct statistical inference on the population mean $\mu_0 = N^{-1} \sum_{i=1}^N y_i$ under the assumed two parametric models \( f(y \mid x; \xi) \) and \( \pi^A(x, y; \theta) \) with the data structure given in \eqref{np.data.structure}.

\subsection{Parameter identifiability} 
Parameter identifiability is a well-known challenge issue in nonignorable missing data problems, regardless of the specific model assumptions \citep{miao2016identifiability,liu2022full, li2023instability}.  Our current investigation with nonignorable non-probability samples is susceptible to the same issue. Before conducting valid statistical inference on the parameters in  \( f(y \mid x; \xi) \) and \( \pi^A(x, y; \theta) \), and ultimately on $\mu_0$,  it is crucial to determine specific conditions under which the parameters are identifiable.
Given that  both $y$ and $x$ are observed in the non-probability sample $\mathcal{S}_A$, we assume that $\xi$ is identifiable.
For simplicity of discussion, we will treat $\xi$ as known in this subsection. Our focus is to determine conditions for identifying $\theta$.

Since the auxiliary variables $x$ are observed in both $\mathcal{S}_A$ and $\mathcal{S}_B$, 
$\pr(x \mid R=1)$ and $\pr(x)$ are identifiable. 
Moreover, $\pr(R=1)$ can be consistently estimated as $n_A / \hat{N}_B$, where $\hat{N}_B = \sum_{i \in \mathcal{S}_B} d_i^B$. 
Consequently, 
$\pr(R=1 \mid x) = \{\pr(x \mid R=1) \pr(R=1)\}/\{\pr(x)\}$
becomes identifiable. It follows from Equation (5) of  Li et al. (2023) that 
\begin{equation}
\label{def.pi.fun}
\pi(x ; \theta, \xi)=
\pr(R=1 \mid x) = \frac{1}{1+\exp \left\{\alpha + x^\T\beta + c(x ; \gamma, \xi)\right\}}
\end{equation}
depends on both set of parameters $\theta$ and $\xi$, where  $c(x; \gamma, \xi) = \log \left\{\mathbb{E}(e^{\gamma y} \mid x, R=1)\right\}$. 
Given the identifiablity of $\pr(R=1 \mid x)$ or equivalently $\pi(x ; \theta, \xi)$, 
the identification of $\theta$ is equivalent to 
identifying $\theta$ in $\alpha + x^\T\beta + c(x ; \gamma, \xi)$.

Let $x = (u^\T, z^\T)$. We call $z$ an instrumental variable if 
\begin{equation*}
\pr(R=1 \mid x, y)=\pr(R=1 \mid u, y)=
\frac{1}{1 + \exp (\alpha + u^\T\beta_1 + \gamma y )}\,,
\end{equation*}
where $\beta_1$ is the vector of components in $\beta$ corresponding to $u$. Note that $\pr(y \mid x, R=1)=\pr(y \mid u,z, R=1)$ depends on $z$ and possibly on $u$. 
The following proposition summarizes two cases in which $\theta$ is identifiable. 

\begin{proposition}\label{prop_identify}
Suppose the participation probability model \eqref{eq:model_prop_score} holds and the probability density/mass function of $y$ given $(x, R = 1)$ is $f(y \mid x, \xi)$.
\begin{itemize}
\item[(a)] The parameters $\theta = (\alpha, \beta^\T, \gamma)^\T$ are identifiable if and only if $\theta$ are identifiable in $\alpha + x^\T\beta + c(x ; \gamma, \xi)$.
\item[(b)] 
If there exists an instrument variable $z$ in $x$, then the parameters $\theta$ are identifiable.
\end{itemize}
\end{proposition}

\subsection{Maximum pseudo likelihood inference for model parameters}

We propose a two-step procedure to estimate the unknown parameters $\xi$ and $\theta$  in \( f(y \mid x; \xi) \) and \( \pi^A(x, y; \theta) \). We assume that the parameters are identifiable. In step 1, 
 we estimate the parameter $\xi$ using the maximum likelihood estimator based on the non-probability sample $\mS_A$ as follows:
 \begin{equation}
 \label{def.hat.xi}
 \hat\xi = \arg\max_{\xi}\sum_{i\in \mS_A} \log f(y_i \mid x_i; \xi).
\end{equation}
In step 2, we estimate the parameter $\theta$ in \( \pi^A(x, y; \theta) \) as follows. Note that the full log-likelihood function based on the complete population dataset $\left\{\left(R_i, x_i\right), i=1,2, \cdots, N\right\}$ is given by
$$
\begin{aligned}
&\sum_{i=1}^N \left[ R_i \log \left\{ \pr(R_i=1 \mid x_i)\right\} + (1-R_i) \log \left\{ \pr(R_i=0 \mid x_i) \right\} \right] \\
=& \sum_{i=1}^N R_i \log \left\{ \frac{\pr(R_i=1 \mid x_i)}{\pr(R_i=0 \mid x_i)} \right\} + \sum_{i=1}^N \log \left\{ \pr(R_i=0 \mid x_i) \right\}.
\end{aligned}
$$
The first term of the full log-likelihood depends solely on the non-probability sample $\mathcal{S}_A$ with $R_i=1$, while the second term represents a population total that can be estimated using the reference probability sample $\mathcal{S}_B$. Thus, we define the pseudo log-likelihood function as 
$$
\begin{aligned}
\ell(\theta, \xi) &= \sum_{i \in \mathcal{S}_A} \log \left\{ \frac{\pr(R_i=1 \mid x_i)}{\pr(R_i=0 \mid x_i)} \right\} + \sum_{i \in \mathcal{S}_B} d_i^B \log \left\{ \pr(R_i=0 \mid x_i) \right\} \\
&= - \sum_{i \in \mathcal{S}_A} \left\{ \alpha + x_i^\T\beta + c(x_i ; \gamma, \xi) \right\} + \sum_{i \in \mathcal{S}_B} d_i^B \left\{ \alpha + x_i^\T\beta + c(x_i ; \gamma, \xi) \right\} \\
&\quad - \sum_{i \in \mathcal{S}_B} d_i^B \log \left[ 1 + \exp \left\{ \alpha + x_i^\T\beta + c(x_i ; \gamma, \xi) \right\} \right],
\end{aligned}
$$
where we have used \eqref{def.pi.fun} in the second equation. 
We propose estimating $\theta$ by
maximizing the pseudo log-likelihood $\ell(\theta, \xi)$ with $\xi$ replaced by $\hat\xi$ obtained in \eqref{def.hat.xi}, that is, 
\begin{equation}
\label{def.hat.theta}
\hat\theta = \arg\max_{\theta}\{\ell(\theta, \hat\xi)\}.
\end{equation}

\subsection{Estimation of the population mean}

We construct three estimators for the population mean $\mu_0=\sum_{i=1}^Ny_i/N$ using IPW, outcome regression, and AIPW techniques. Let $\hat\theta$ be obtained from (\ref{def.hat.theta}). The participation probabilities are estimated by $\pi^A(x_i, y_i; \hat\theta)$ for $i\in \mS_A$. The IPW estimator of $\mu_0$ is computed as 
\[
\hat\mu_{IPW} = \frac{1}{\hat N_A} \sum_{i \in \mS_A} \frac{y_i}{\pi^A(x_i, y_i; \hat\theta)},
\]
where \(\hat N_A=\sum_{i \in \mS_A} 1/\pi^A(x_i, y_i; \hat\theta)\).

To construct the regression-based prediction estimator of $\mu_0$, it is essential to determine the probability density/mass function of $y$ given $x$ and to estimate $\e(y\mid x)$.
In Section 1 of the supplementary material, we show the following proposition. 
\begin{proposition}
\label{prop.conditional}
Suppose that the participation probability model \eqref{eq:model_prop_score} holds, and the probability density/mass function of $y$ given $(x, R = 1)$ is specified as $f(y \mid x, \xi)$. We have 
\begin{itemize}
\item[(a)] the conditional probability density/mass function of $y$ given $x$
is 
\begin{equation*}
\pr(y \mid x)=\pi(x ; \theta, \xi) f(y \mid x, \xi) 
+\{1-\pi(x ; \theta, \xi) \} f(y \mid x, \xi)  \exp\{\gamma y - c(x ; \gamma, \xi)\};
%\label{form.con.pdf} % Yang Liu: not cited
\end{equation*}

\item[(b)] the conditional expecation of $y$ given $x$ is 
\begin{equation*}
m(x ; \theta, \xi) =
\e(y\mid x)=
\pi(x ; \theta, \xi) \nabla_\gamma c(x ; 0, \xi) + \{1-\pi(x ; \theta, \xi)\} \nabla_\gamma c(x ; \gamma, \xi),
%\label{form.ce} % Yang Liu: not cited
\end{equation*}
where $\nabla_\gamma c(x ; \gamma, \xi)$ denotes the first partial derivative of $c(x ; \gamma, \xi)$
with respect to $\gamma$.
\end{itemize}
 
\end{proposition}

Using the estimators $\hat{\theta}$ and $\hat{\xi}$  obtained from (\ref{def.hat.xi}) and (\ref{def.hat.theta}) and the observed $x$ variables in the reference probability sample $\mS_B$, the regression-based prediction estimator is constructed as 
$$
\hat{\mu}_{REG} = \frac{1}{\hat{N}_B} \sum_{i \in \mathcal{S}_B} d_i^B m(x_i ; \hat{\theta}, \hat{\xi}) \,,
$$
where $\hat{N}_B = \sum_{i \in \mathcal{S}_B} d_i^B$. 

If the complete population auxiliary information $\{(x_i, i=1,2, \cdots, N\}$ was available in addition to the non-probability sample dataset, the standard AIPW estimator of $\mu_0$ can be expressed as
$$
\frac{1}{N} \sum_{i=1}^N \frac{R_i\{y_i - m(x_i ; \hat{\theta}, \hat{\xi})\}}{\pi^A(x_i, y_i ; \hat{\theta})} + \frac{1}{N} \sum_{i=1}^N m(x_i ; \hat{\theta}, \hat{\xi}).
$$
Under the current setting of two samples $\mS_A$ and $\mS_B$,  our proposed AIPW estimator of $\mu_0$ is given by 
$$
\hat\mu_{AIPW} = \frac{1}{\hat N_A} \sum_{i \in \mS_A} \frac{y_i - m(x_i; \hat\theta, \hat\xi)}{\pi^A(x_i, y_i; \hat\theta)} + \frac{1}{\hat N_B} \sum_{i \in \mS_B} d_i^B m(x_i; \hat\theta, \hat\xi).
$$

\begin{remark}
When the participation mechanism is ignorable, the AIPW estimator $\hat\mu_{AIPW}$ aligns with the double robust estimator proposed by \cite{chen2020doubly}. However, the double robustness property does not hold when the participation mechanism is nonignorable, as the conditional mean $m(x; \theta, \xi)$ is intricately dependent on both $\pi(x; \theta, \xi)$ and $f(y \mid x; \xi)$. 
\end{remark}

\subsection{Asymptotic properties and variance estimation\label{section.variance}}

%%Before presenting the asymptotic behaviors of the three estimators ($\hat\mu_{REG}$, $\hat\mu_{IPW}$, and $\hat\mu_{AIPW}$), we introduce the necessary notation. 

Let $\theta_0= (\alpha_0, \beta_0^\T,\gamma_0)^\T$ be the true values of $\theta= (\alpha, \beta^\T,\gamma)^\T$ and $\xi_0$ be the value of $\xi$ as the solution to the ``census equation''
$$
\sum_{i=1}^N \pi^A(x_i, y_i;\theta_0)
\nabla_\xi\log\{f(y_i \mid x_i; \xi)\} = 0.
$$
Let
$\pi^A_i = \pi^A(x_i, y_i; \theta_0)$,
$\pi_i = \pi(x_i; \theta_0, \xi_0)$,
$h^A_i =  (1, x_i^\T, y_i)^\T$,  
$h_i =
(1, x_i^\T, \nabla_\gamma c(x; \gamma_0, \xi_0) )^\T$,  
and $m_i = m(x_i; \theta_0, \xi_0)$, $i=1,\dots,N$. 
Let $ \hbar_N = N^{-1} \sum_{i=1}^N (y_i - m_i)$.  
We further define the following population-level matrices and vectors which are used for the expressions of asymptotic variances:
\begin{align*}
V_{12} =&
\frac{1}{N}
\sum_{i=1}^N
(1-\pi^A_i) (y_i - \mu_0) (h_i^A)^\T,
\quad
V_{22}=
- \frac{1}{N}
\sum_{i=1}^N
\pi_i
(1 - \pi_i)
h_i^{\otimes2},
\\
V_{23} =&
- \frac{1}{N}
\sum_{i=1}^N
\pi_i (1 - \pi_i)
h_i \{\nabla_{\xi} c(x_i; \gamma_0, \xi_0)\}^\T,
\quad
V_{33}= \frac{1}{N}
\sum_{i=1}^N \pi^A_i
\nabla^2_{\xi\xi^\T}
\log\{ f(y_i\mid x_i; \xi_0) \},
\\ 
V_{12e} =& 
\frac{1}{N}
\sum_{i=1}^N
\{\nabla_\theta m(x_i; \theta_0, \xi_0)\}^\T,
\quad
V_{13e} = 
\frac{1}{N}
\sum_{i=1}^N
\{\nabla_\xi m(x_i; \theta_0, \xi_0)\}^\T,
\\
V_{12a} =& 
\frac{1}{N}
\sum_{i=1}^N
(1-\pi^A_i)
(y_i - m_i - \hbar_N) (h_i^A)^\T, 
%\\
%W_{11} =& \frac{1}{N} \sum_{i=1}^N
%\frac{1 - \pi^A_i}
%{\pi^A_i}\cdot
%(y_i -\mu_0)^2,
%\quad
%W_{21} = %W_{12}^\T =
%- \frac{1}{N} \sum_{i=1}^N
%(1 - \pi_i)
%(y_i - \mu_0)
%h_i,
%\\
%W_{31} =&
%\frac{1}{N} \sum_{i=1}^N
%(1 - \pi^A_i)
%(y_i - \mu_0)
%\nabla_\xi \log f(y_i\mid x_i; \xi_0),
%\\
%W_{22} =& \frac{1}{N} \sum_{i=1}^N
%\pi_i (1 - \pi_i) h_i^{\otimes2} +
%\mathbb{V}_B\left(\frac{1}{\sqrt N}
%\sum_{i\in \mS_B} d_i^B\pi_i
%h_i\right),
%\\
%W_{32} =& %W_{23}^\T  = 
%-\frac{1}{N} \sum_{i=1}^N
%\pi^A_i (1 - \pi_i) 
%\nabla_{\xi}
%\log\{f(y_i\mid x_i; \xi_0)\} h_i^\T,
%\\
%%W_{23}  = & W_{32}^\T = -
%%\frac{1}{N} \sum_{i=1}^N
%%\pi^A_i (1 - \pi_i) h_i
%%\{\nabla_{\xi}
%%\log\{f(y_i\mid x_i; \xi_0)\}^\T,
%%\\
%W_{33} =& \frac{1}{N} \sum_{i=1}^N
%\pi^A_i (1 - \pi^A_i)
%\{\nabla_{\xi} \log\{f(y_i\mid x_i; \xi_0)\}^{\otimes2},
%\\
%%W_{11e} =& \mathbb{V}_B\left(
%%\frac{1}{\sqrt N} \sum_{i\in \mS_B}
%%d_i^B (m_i - \mu_0)
%%\right),
%%\quad
%%W_{31e} = W_{13e}^\T = 0_{d_\xi\times1},
%%\\
%W_{21e} =& %W_{12e}^\T =
%\frac{1}{N} \sum_{i=1}^N
%\left\{1 - (d_i^B)^{-1}\right\} \pi_i h_i (m_i - \mu_0),
\end{align*} 
where $A^{\otimes2} = AA^\T$ for a vector $A$, $\nabla^2$ is the second-order partial derivative operator with respect to the subscript parameters.  
Let $\mathbb{V}_B(\cdot)$ represent the design-based variance under the probability sampling design for $\mS_B$.

\begin{theorem}\label{thm-asy}
Suppose the conditions in Proposition \ref{prop_identify} and 
the regularity conditions C1--C7 in Section 2 of the supplementary material are satisfied. 
As $N\to\infty$, we have\\ (a)
$\sqrt N (\hat \mu_{IPW} - \mu_0)/\sigma_{IPW}\overset{d}{\rightarrow}
N(0, 1)$, where  
$$
\begin{aligned} 
 \sigma_{IPW}^2=&
\frac{1}{N}
\sum_{i=1}^N \pi_i^A(1 - \pi_i^A)
\left[
\frac{y_i - \mu_0}
{\pi^A_i}  + V_{12}V_{22}^{-1} h_i
+V_{12}V_{22}^{-1}V_{23}V_{33}^{-1} 
\nabla_\xi \log\{f(y_i\mid x_i; \xi_0)\} \right]^2
\\
 &\qquad\qquad +
\frac{1}{N} \mathbb{V}_B
\left(\sum_{i\in \mS_B} d_i^B \pi_iV_{12}V_{22}^{-1} h_i \right).
\end{aligned}
$$ 
(b) $\sqrt N (\hat \mu_{REG} - \mu_0)/\sigma_{REG}\overset{d}{\rightarrow}
N(0, 1)$, where 
$$
\begin{aligned}
%\label{eq:sigma_AIPW}\notag
\sigma_{REG}^2=&
\frac{1}{N}
\sum_{i=1}^N \pi_i^A(1 - \pi_i^A)
\left[
 V_{12e}V_{22}^{-1} h_i + 
(V_{12e}V_{22}^{-1}V_{23} -V_{13e} )V_{33}^{-1}
\nabla_\xi \log\{f(y_i\mid x_i; \xi_0)\} \right]^2
\\
 &\qquad\qquad +
\frac{1}{N} \mathbb{V}_B
\left(\sum_{i\in \mS_B}  
d_i^B \left(m_i -\mu_0 - V_{12e}V_{22}^{-1}\pi_i h_i\right)\right).
\end{aligned}
$$
 (c) $\sqrt N (\hat \mu_{AIPW} - \mu_0)/\sigma_{AIPW}\overset{d}{\rightarrow}
N(0, 1)$, where
$$
\begin{aligned}
%\label{eq:sigma_AIPW}\notag
\sigma_{AIPW}^2=&
\frac{1}{N}
\sum_{i=1}^N \pi_i^A(1 - \pi_i^A)
\bigg[
\frac{y_i - m_i - \hbar_N}
{\pi^A_i} +
V_{12a}V_{22}^{-1} h_i 
\\
&\qquad\qquad\qquad\qquad\qquad +V_{12a} V_{22}^{-1}V_{23}V_{33}^{-1}\cdot
\nabla_\xi \log\{f(y_i\mid x_i; \xi_0)\}
\bigg]^2
\\
 &\qquad\qquad +
\frac{1}{N} \mathbb{V}_B
\left(\sum_{i\in \mS_B} d_i^B
\left(m_i
- \frac{1}{N}
\sum_{j=1}^N
m_j -\pi_i V_{12a}V_{22}^{-1}
h_i
\right)\right).
\end{aligned}
$$
\end{theorem}

The asymptotic variance formulas presented in Theorem \ref{thm-asy} can be used to construct plug-in variance estimators for the three point estimators for the population mean. The elements involved for the plug-ins are matrices and vectors which can be expressed as either
\[
N^{-1}\sum_{i=1}^N g(x_i, y_i; \mu_0, \theta_0, \xi_0) \quad \text{or} \quad \mathbb{V}_B\left(N^{-1/2}\sum_{i\in \mS_B} d_i^B g(x_i; \mu_0, \theta_0, \xi_0)\right)
\]
for some function \(g\). Since both response and auxiliary variables are observed in the non-probability sample \(\mS_A\), a consistent estimator of the first quantity is
\[
\frac{1}{\hat{N}_A}\sum_{i\in \mS_A} \frac{g(x_i, y_i; \hat\mu, \hat\theta, \hat\xi)}{\pi^A(x_i, y_i; \hat{\theta})}.
\]
Based on the probability sample \(\mS_B\), the design-based estimator of the second quantity is given by
\begin{equation}\label{eq:prob-design}
\frac{1}{\hat{N}_B}\sum_{i\in \mS_B}\sum_{j\in \mS_B} \frac{\pi_{ij}^B - \pi_i^B \pi_j^B}{\pi_{ij}^B} \frac{g(x_i; \hat\mu, \hat\theta, \hat\xi)}{\pi_i^B} \frac{\{g(x_j; \hat\mu, \hat\theta, \hat\xi)\}^\T}{\pi_j^B},
\end{equation}
where \(\pi_i^B\) and \(\pi_{ij}^B\) are the first and second order inclusion probabilities for the probability sample \(\mS_B\). For certain sampling designs, computing the second order inclusion probabilities \(\pi_{ij}^B\) can be theoretically challenging and computationally complex. In such cases, approximate estimators for the design-based variance from the survey sampling literature, such as those proposed by \cite{hajek1964asymptotic} and \cite{berger2004simple},  can be used.

\section{Simulation Studies\label{Section3}}

In this section, we report results from simulation studies to illustrate the finite-sample performance of the maximum pseudo-likelihood estimator  of the model parameters $\theta$ as defined in \eqref{def.hat.theta}, as well as the regression, IPW, and AIPW estimators of the population mean $\mu_0$.

\subsection{Simulation setup}

We consider a finite population of size \( N = 20,000 \). For unit $i$ \( (i = 1, 2, \ldots, N) \), 
the auxiliary variables are \( x_i = (u_i^\T, z_i)^\T \), 
where \( u_i = (u_{i1}, u_{i2})^\T \) follows a standard bivariate normal distribution, and \( z_i \)  is uniformly distributed on [0,3]. These variables are independent. We assume that \( z_i \) serves as an instrumental variable, influencing the conditional distribution of the response variable without affecting the participation mechanism.

To mimic the real example to be presented in Section 4, we consider a binary response $y$.
Let $Bern(p)$ denote the Bernoulli distribution with the success probability of $p$. 
%\red{Tentatively, we omit the subscript $i$ and focus on a specific auxiliary variable vector, denoted as $x = (u_1, u_2, z)^\T$.} % added by Yang Liu
The generation of the response $y$ and the non-probability sample \(\mS_A\)
is based on 
two parametric models:
\begin{equation}
\label{eq:sim-response}
y\mid(x,R=1)\sim Bern\big(c_1(x)\big)\mbox{ with }
c_1(x)=\frac{\exp(-1.8 + 1.2u_{1}+ 1.2 u_2+z)}{1 + \exp(-1.8+ 1.2u_{1}+ 1.2 u_2+z)},
\end{equation}
and 
\begin{equation}\label{eq:sim-nonprob}
\pr(R=1 \mid x, y)= \frac{1}{1 + \exp (\alpha+\beta_1 u_1+ \beta_2 u_2+\gamma y)},\quad
(\beta_1, \beta_2) = (-0.7, 1.5).
\end{equation}
We consider two scenarios for the parameter \(\gamma\): 0.8 and \(-0.8\). Given these specified parameters, the true value of \(\alpha\) is adjusted to ensure that  the expected size of the non-probability sample is either 500 or 2000.
Table \ref{para.setting} summarizes the parameter settings for the simulation studies. 

\medskip
\begin{table}[!htt]
\begin{center} 
\caption{Simulation settings of the model parameters ($\alpha$ and $\gamma$), the expected non-probability sample size  $E(n_A)$, and the corresponding population mean $\mu_0$.}\label{para.setting}
\setlength{\tabcolsep}{0.5pt}{
\begin{tabular*}{\textwidth}{@{\extracolsep{\fill}}rrrr@{\extracolsep{\fill}}}
\toprule  
$\alpha$ & $\gamma$ & $E(n_A)$ &  $\mu_0$
\\
\midrule
 
4.5&0.8 &500&0.58
\\                    

2.7 &0.8 &2000&0.57
\\
5.1&-0.8 &500&0.34
\\                    

3.3 &-0.8 &2000&0.35
\\
 
\bottomrule
\end{tabular*}} 
\end{center}
\end{table}

Under \eqref{eq:sim-response} and \eqref{eq:sim-nonprob}, by Proposition \ref{prop.conditional} (a), it follows that 
\begin{equation}
\label{eq:sim-y-con-x}
y\mid x\sim \pi(x) Bern\big(c_1(x)\big)+\{1-\pi(x)\}Bern\big(c_0(x)\big),
\end{equation}
where $c_1(x)$ is given in \eqref{eq:sim-response}, 
$$
c_{0}(x)=
\frac{\exp(\gamma-1.8 + 1.2u_{1}+ 1.2u_2+z)}{1 + \exp(\gamma-1.8 + 1.2u_{1}+ 1.2u_2+z)},
$$
and 
$$
\pi(x)=\frac{1}{1 + \exp \{\alpha-0.7u_1+1.5 u_2+c(x)\}}
~~\mbox{ with }~~
c(x) = \log\left\{\frac{1 - c_{1}(x)}{1 - c_{0}(x)}\right\}.
$$

For each parameter setting in Table \ref{para.setting}, and given the auxiliary variables 
$x_i$, we generate the response 
$y_i$ for unit $i$ \( (i = 1, 2, \ldots, N) \) using model \eqref{eq:sim-y-con-x}. 
Consequently, the resulting finite population consists of  $N$ units, each with a response 
 $y_i$ and corresponding auxiliary variables 
$x_i$ for $i=1,\ldots,N$. 
Using the generated finite population, we replicate the simulation 500 times. In each repetition, a non-probability sample \(\mS_A\) is drawn using Poisson sampling with participation probabilities specified by \eqref{eq:sim-nonprob}, and a probability sample 
\(\mS_B\), with size 
\(n_B\) of either 1000 or 2000, drawn by simple random sampling without replacement.
 For sample \(\mS_A\), both the observed \(x_i\) and \(y_i\) are retained, whereas for  sample \(\mS_B\), only observed \(x_i\) are kept. The survey weights for the reference sample are given by  $d_i^B=N/n_B$ for unit $i\in \mS_B$.

\subsection{Comparison of methods for estimating participation probabilities}  

We evaluate the performance of the proposed pseudo-likelihood estimation method
for estimating $\theta=(\alpha,\beta_1,\beta_2,\gamma)^\T$ in the participation probability model \eqref{eq:sim-nonprob}, and compare it with the calibration method used in \cite{kim2023empirical}. Although the calibration method was originally designed for scenarios where the auxiliary variables are available for the entire finite population, it can be easily adapted to our current setup with a reference probability sample. Additional technical details are given in the Appendix.

We summarize the results in terms of relative bias (\%RB) and relative root mean squared error (RRMSE) in 
Table \ref{tab:simBinaryPointEstTheta}. 
For any scalar-valued parameter $\zeta$, these two assessment criteria are defined as follows: 
$$
{\rm \%RB} = \frac{1}{B} \sum_{b=1}^{B} \frac{\hat \zeta_b - \zeta_0}{\zeta_0} \times 100
\quad \text{and} \quad
{\rm RRMSE} = \left\{\frac{1}{B} \sum_{b=1}^{B} \frac{(\hat \zeta_b - \zeta_0)^2}{\zeta_0^2}\right\}^{1/2},
$$
where $\zeta_0$ is the true value of $\zeta$, $\hat \zeta_b$ denotes the estimate from the $b$th simulated sample, and $B$ is the number of replications. 
Note that during the implementation of the calibration method, we encountered multiple solutions for  Equation  \eqref{eq:cal-theta} in some cases. The last two columns of Table \ref{tab:simBinaryPointEstTheta} show the number of cases where multiple roots were absent for the calibration method and the pseudo-likelihood method, respectively. The reported \%RB and RRMSE values for each method are based on the cases where  no multiple solutions were found.

\medskip

\begin{table}[!htt]
\begin{center} 
\caption{Number of cases without multiple roots (NMR), relative biases (in percentage: \%RB), and relative root mean squared errors (RRMSE) for the calibration estimator (CAL) and the maximum pseudo-likelihood estimator (PL) of  $(\alpha, \beta_1, \beta_2, \gamma)^\T$.
}\label{tab:simBinaryPointEstTheta}
\setlength{\tabcolsep}{1pt}{
\begin{tabular*}{\textwidth}{@{\extracolsep{\fill}}rrrrrrrrrrrr@{\extracolsep{\fill}}}
\toprule 
&&\multicolumn{2}{@{}c@{}}{$\alpha$}&\multicolumn{2}{@{}c@{}}{$\beta_1$}&\multicolumn{2}{@{}c@{}}{$\beta_2$}&\multicolumn{2}{@{}c@{}}{$\gamma$}&\multicolumn{2}{@{}c@{}}{NMR}
\\ \cline{3-4}\cline{5-6}\cline{7-8}\cline{9-10}\cline{11-12}
$\big(E(n_A), n_B\big)$
& &CAL&PL&CAL&PL&CAL&PL&CAL&PL&CAL&PL\\
\midrule
\multicolumn{12}{@{}c@{}}{$\gamma=0.8$}
\\
 (500, 1000)
%&NMR &454&500&489&500&490&500&450& 500
%\\

&\%RB &-2.00 &  0.34 & 11.06 &  2.09 & 15.89  & 2.09  & 4.05 & -1.92&452&500
                                                           
\\
&RRMSE&0.26  & 0.10 &  0.51 &  0.25 &  0.31  & 0.15 &  2.01 &  0.85
                 
\\  
(500, 2000)
%&NMR &466&500&491&500&493&500&463& 500
%\\
&\%RB &-0.93  & 0.10 &  6.77&  -0.48 & 13.29  & 1.15 & -7.78  &-0.44&465&500

\\
&RRMSE&0.19  & 0.07  & 0.48 &  0.18 &  0.27 &  0.11  & 1.72  & 0.61

\\
(2000, 1000)
%&NMR &498&500&500&500&500&500&498& 500
%\\
&\%RB &-2.96 & -0.29 &  3.81  & 1.21 &  2.81 &  1.39 & 10.65 &  0.50&498&500
        
\\
&RRMSE&0.16  & 0.12 &  0.23 &  0.21  & 0.13  & 0.12 &  0.88 &  0.69                                                     
\\

(2000, 2000)
%&NMR &499&500&500&500&500&500&499& 500
%\\
&\%RB &-2.48 &  0.39  & 4.34 & -0.84 &  3.79 &  1.18  & 8.21 & -3.36&499&500 

\\
&RRMSE&0.18 &  0.09 &  0.24 &  0.15 &  0.12 &  0.09  & 0.92  & 0.51
                                                         
\\
\multicolumn{12}{@{}c@{}}{$\gamma=-0.8$}
\\
(500, 1000)
%&NMR &496&500&496&500&496&500&454& 500
%\\
&\%RB &-2.65  & 0.51 & 15.07  & 2.33  & 8.85 &  1.83 &-12.78  & 8.46&454& 500
\\
&RRMSE&0.25 &  0.06 &  0.41 &  0.22 &  0.25  & 0.13 &  2.28  & 0.89

\\ 
(500, 2000)
%&NMR &497&500&500&500&499&500&464& 500
%\\
&\%RB &-3.52 &  0.03 & 13.12  & 0.46 &  7.18 &  0.53 & -9.27  & 3.69&464& 500
\\
&RRMSE&0.28  & 0.04 &  0.43  & 0.16 &  0.23  & 0.09  & 2.49  & 0.66
               
\\
 (2000, 1000)
%&NMR &500&500&500&500&500&500&495& 500
%\\
&\%RB &-0.04  & 0.14 &  1.97  & 1.41 &  3.23  & 1.15 &  8.46  & 3.79&495& 500
 \\
&RRMSE&0.07 &  0.07 &  0.20  & 0.18 &  0.12 &  0.11 &  0.89 &  0.69  
\\
(2000, 2000)
%&NMR &500&500&500&500&500&500&496& 500
%\\
&\%RB &0.24 &  0.15  & 2.89 & -0.11 &  3.72 &  1.01&  10.50 &  3.19&496& 500
\\
&RRMSE&0.06  & 0.05 &  0.19  & 0.14 &  0.12 &  0.08  & 0.87  & 0.52
\\                                           
\bottomrule
\end{tabular*}} 
\end{center}
\end{table}

From simulation results reported in Table \ref{tab:simBinaryPointEstTheta}, 
we have the following observations. 
(1) The calibration method faces the issue of multiple roots, especially for small $E(n_A)$. In contrast, the pseudo-likelihood method does not encounter this issue.
 (2) Even after excluding cases with multiple roots, the calibration estimators consistently show larger \%RB and RRMSE compared to the maximum pseudo-likelihood estimators for all elements of $\theta$.
(3)  The \%RBs of maximum pseudo-likelihood estimators
 are generally negligible, except for $\gamma=-0.8$ and $(E(n_A), n_B)=(500,1000)$, where the  \%RB for estimating $\gamma$ is 8.46\%.
 The \%RB decreases significantly as $E(n_A)$ or $n_B$ increases.

\subsection{Comparison of point estimators for the population mean}

We evaluate the performance of point estimators for the population mean $\mu_0 = N^{-1}\sum_{i=1}^N y_i$. The estimators under consideration include the proposed regression-based prediction estimator ($\hat\mu_{REG}$), the IPW estimator ($\hat\mu_{IPW}$), and the AIPW estimator ($\hat\mu_{AIPW}$). We also include three alternative estimators ($\hat\mu_{REG2}$, $\hat\mu_{IPW2}$, and $\hat\mu_{DR2}$) introduced by \cite{chen2020doubly} under the ignorable participation mechanism, as well as the naive sample mean $\bar y_A = n_A^{-1}\sum_{i\in \mS_A} y_i$ for comparisons. In addition, we extend \cite{kim2023empirical}'s empirical likelihood (EL) method to our current setup; technical details are provided in the Appendix. The corresponding EL estimator of the population mean is denoted as $\hat \mu_{EL}$.

\medskip
\begin{table}[!htt]
\begin{center} 
\caption{Relative biases (in percentage: \%RB) and
relative root mean squared errors (RRMSEs) 
of the eight estimators of the population mean $\mu_0$.
}\label{tab:simBinaryPointEst}
\setlength{\tabcolsep}{1pt}{
\begin{tabular*}{\textwidth}{@{\extracolsep{\fill}}rrrrrrrrrr@{\extracolsep{\fill}}}
\toprule 
$\big(E(n_A), n_B\big)$
& &$\bar y_A$&
$\hat \mu_{REG2}$&
$\hat \mu_{IPW2}$&
$\hat \mu_{DR2}$&
$\hat \mu_{REG}$&
$\hat \mu_{IPW}$&
$\hat \mu_{AIPW}$&
$\hat \mu_{EL}$
\\
\midrule
\multicolumn{10}{@{}c@{}}{$\gamma=0.8$}
\\   
 (500, 1000)
&\%RB &-50.64 &-18.74 &-21.97 & -20.46 & 0.38& -2.02 &-1.28  & -2.14
\\
&RRMSE&0.51 &  0.20 &  0.26 &   0.23&  0.17&  0.22 & 0.20 &   0.34
\\   
(500, 2000)
&\%RB &-50.89 &-18.93 &-21.50 & -20.39 & 0.71& -1.02 &-0.71 &  -3.72
\\
&RRMSE&0.51 &  0.20  & 0.26   & 0.23 & 0.13&  0.18 & 0.15   & 0.31
\\                          
(2000, 1000)
&\%RB &-44.50& -18.63& -19.82 & -19.30&  0.28 &-0.16& -0.14 &   0.09
\\
&RRMSE&0.45  & 0.19 &  0.21  &  0.20 & 0.13 & 0.15&  0.13   & 0.17
\\   
(2000, 2000)
&\%RB &-44.59& -18.80& -20.02&  -19.68 &-0.40& -1.05 &-1.12  & -1.29
\\
&RRMSE& 0.45  & 0.19 &  0.21 &   0.20 & 0.10 & 0.12 & 0.11  &  0.16
\\
\multicolumn{10}{@{}c@{}}{$\gamma=-0.8$}
\\   
(500, 1000)
&\%RB &16.18 & 35.79 & 36.49 &  34.63& -0.21 &-1.24 &-0.17 &   7.30
\\
&RRMSE& 0.17 &  0.37 &  0.42 &   0.38 & 0.29 & 0.34 & 0.30  &  0.51
\\  
(500, 2000) 
&\%RB &16.50 & 35.96  &37.72&   34.84&  0.81&  0.56  &0.47 &   6.97
\\
&RRMSE& 0.18 &  0.37  & 0.43  &  0.38 & 0.24 & 0.30&  0.26 &   0.55
\\  
 (2000, 1000)
&\%RB & 16.89 & 33.05 & 33.56  & 32.07 & 1.16 &-0.14  &0.35   & 1.13
\\
&RRMSE&0.17 &  0.33  & 0.35    &0.33 & 0.21  &0.24 & 0.22  &  0.26
\\  
(2000, 2000) 
&\%RB &16.86 & 33.23  &34.36 &  32.08 & 1.12&  0.49 & 0.01  &  1.32
\\
&RRMSE&0.17 &  0.34  & 0.36  &  0.33 & 0.16&  0.19 & 0.17   & 0.24
\\
\bottomrule
\end{tabular*}} 
\end{center}
\end{table}

Simulation results on \%RBs and RRMSEs for the eight estimators are presented in Table \ref{tab:simBinaryPointEst}.
It can be seen that the naive sample mean $\bar y_A$ and the estimators ($\hat \mu_{REG2}$, $\hat \mu_{IPW2}$, and $\hat \mu_{DR2}$) assuming an ignorable participation mechanism exhibit significant biases. These biases are negative when $\gamma=0.8$ and positive when $\gamma=-0.8$.
This pattern likely arises due to the non-representative nature of the non-probability sample. 
Specifically, the propensity to participate in the survey varies based on the respondent's response value (one or zero) and the sign of $\gamma$. When $\gamma=0.8$,  individuals with a response value of one are less likely to participate, leading to 
their underrepresentation in the sample. Conversely, when $\gamma=-0.8$, these respondents are more likely to participate, resulting in their overrepresentation. When considering a nonignorable participation mechanism, the biases of the proposed estimators ($\hat \mu_{REG}$, $\hat \mu_{IPW}$, and $\hat \mu_{AIPW}$) as well as the \cite{kim2023empirical}’s EL estimator ($\hat \mu_{EL}$) are significantly mitigated. The three proposed estimators consistently exhibit smaller  RRMSEs than \cite{kim2023empirical}’s EL estimator. 
Among the three proposed estimators, the regression-based prediction estimator $\hat \mu_{REG}$ has the smallest RRMSE across all settings.

\subsection{Simulation results on variance estimation}

The reference probability sample is selected by simple random sampling without replacement with first- and second-order inclusion probabilities given by $\pi_i^B = n_B/N$ and $\pi_{ij}^B = n_B(n_B-1)/\{N(N-1)\}$. 
The plug-in variance estimators for the three proposed estimators of the population mean $\mu_0$ are computed based on 
the techniques discussed in Section \ref{section.variance}.
Let $\hat\mu_b$ and $\hat v_{b}$
be the point estimator $\hat\mu$
and the corresponding plug-in variance estimator $\hat{v}$
computed from the $b$th simulation sample. 
The simulated standard error (SE) and standard deviation (SD) are computed as 
$$
{\rm SE}=\frac{1}{B}
\sum_{b=1}^{B} 
{\hat v_b}^{1/2}~~\mbox{ and }
~~
{\rm SD}=\left\{\frac{1}{B-1} \sum_{b=1}^{B} \Big(\hat\mu_b - \frac{1}{B}\sum_{l=1}^B\hat\mu_l\Big)^2\right\}^{1/2} \,.
$$

\begin{table}[!htt]
\begin{center} 
\caption{Simulated standard deviations (SDs) and standard errors (SEs) 
of the proposed regression, IPW, and AIPW estimators of the population mean $\mu_0$.
}\label{tab:sim_seBinary_mu0}
\setlength{\tabcolsep}{1pt}{
\begin{tabular*}{\textwidth}{@{\extracolsep{\fill}}rrrrrrrr@{\extracolsep{\fill}}}
\toprule 
&&
\multicolumn{3}{@{}c@{}}{$\gamma=0.8$} &
\multicolumn{3}{@{}c@{}}{$\gamma=-0.8$}

\\\cline{3-5}\cline{6-8}
$\big(E(n_A), n_B\big)$
& &$\hat \mu_{REG}$&
$\hat \mu_{IPW}$&
$\hat \mu_{AIPW}$&
$\hat \mu_{REG}$&
$\hat \mu_{IPW}$&
$\hat \mu_{AIPW}$
\\
\midrule
\\
(500, 1000)
&SD   &0.101& 0.128 &0.115 & 0.098 &0.116& 0.101 
\\
&SE   &0.101& 0.123 & 0.108 &0.098 &0.117 & 0.101  
\\ 

(500, 2000)
&SD &0.078 &0.104 &0.088 & 0.081& 0.101& 0.087 
\\   
&SE &0.078& 0.100 & 0.085& 0.078 &0.095 & 0.084  
\\ 

(2000, 1000)
&SD &0.073 &0.088& 0.077& 0.074 &0.083& 0.076 

\\
&SE &0.075& 0.090 & 0.078 &0.074 &0.086 & 0.077
\\ 

(2000, 2000)
&SD &0.058& 0.069& 0.061& 0.056& 0.064& 0.058  
\\
&SE &0.054 &0.066 & 0.057 &0.054& 0.064 & 0.058
\\ 
\bottomrule
\end{tabular*}} 
\end{center}
\end{table}

Table \ref{tab:sim_seBinary_mu0} presents the simulated SDs and SEs for the three proposed estimators of $\mu_0$.
The results show that the SEs are consistently close to the SDs for all cases considered in the simulation, with the largest absolute relative bias being 6.9\% ($|0.054-0.058|/0.058$). Among the three estimators of $\mu_0$, the regression-based prediction estimator performs the best in terms of both SD and SE, followed by the AIPW estimator. The IPW estimator is less stable compared to the other two estimators.

%The results in Table \ref{tab:sim_seBinary} indicates that the SEs of the regression estimator are slightly higher than the corresponding SDs, while the SEs for both the IPW and AIPW estimators are slightly lower than the SDs. Overall, the approximation of SEs to SDs improves as either $E(n_A)$ or $n_B$ increases.

The plug-in variance estimator along with the associated point estimator can be used to construct a Wald-type confidence interval for $\mu_0$ using normal approximations. The simulated coverage probabilities and average lengths (in parentheses) of 95\% confidence intervals are presented in Table \ref{tab:sim_cpBinary_mu0}.  The results show that all three Wald-type confidence intervals of $\mu_0$ have coverage probabilities in the range of  90.4\% $\sim$ 92.2\% when $E(n_A) = 500$, which are lower than the nominal value 95\%. 
As $E(n_A)$ increases to 2000, the coverage probabilities become very close to 95\%.

\medskip

\begin{table}[!htt]
\begin{center} 
\caption{Simulated coverage probabilities (in percentage) and average lengths (in parentheses) of three Wald-type confidence intervals for $\mu_0$.
}\label{tab:sim_cpBinary_mu0}
\setlength{\tabcolsep}{1pt}{
\begin{tabular*}{\textwidth}{@{\extracolsep{\fill}}rcccccc@{\extracolsep{\fill}}}
\toprule 
& 
\multicolumn{3}{@{}c@{}}{$\gamma=0.8$} &
\multicolumn{3}{@{}c@{}}{$\gamma=-0.8$}

\\\cline{2-4}\cline{5-7}
$\big(E(n_A), n_B\big)$
&$\hat \mu_{REG}$&
$\hat \mu_{IPW}$&
$\hat \mu_{AIPW}$&
$\hat \mu_{REG}$&
$\hat \mu_{IPW}$&
$\hat \mu_{AIPW}$
\\
\midrule 
\\
(500, 1000) &91.6 (0.39) &91.4 (0.48) &91.4 (0.42) &91.2 (0.38) &90.6 (0.46) &92.2 (0.40)
\\                    
      
(500, 2000)  &91.8 (0.31) &90.4 (0.39) &91.8 (0.33) &91.6 (0.30) &92.0 (0.37) &91.8 (0.33)
\\         
      
(2000, 1000) &94.2 (0.29) &94.0 (0.35) &94.4 (0.31) &95.0 (0.29) &95.4 (0.34) &94.6 (0.30)
\\
       
(2000, 2000)  &92.8 (0.21) &94.4 (0.26) &94.2 (0.23) &94.4 (0.21) &94.0 (0.25) &94.4 (0.23)
\\

\bottomrule
\end{tabular*}} 
\end{center}
\end{table}

%\medskip
%\begin{table}[!htt]
%\begin{center} 
%\caption{Simulated coverage probabilities (in percentage) and average lengths (in parentheses) of three Wald-type confidence intervals for $\log(\mu_0)$.
%}\label{tab:sim_cpBinary}
%\setlength{\tabcolsep}{1pt}{
%\begin{tabular*}{\textwidth}{@{\extracolsep{\fill}}rcccccc@{\extracolsep{\fill}}}
%\toprule 
%& 
%\multicolumn{3}{@{}c@{}}{$\gamma=0.8$} &
%\multicolumn{3}{@{}c@{}}{$\gamma=-0.8$}
%
%\\\cline{2-4}\cline{5-7}
%$\big(E(n_A), n_B\big)$
%&$\hat \mu_{REG}$&
%$\hat \mu_{IPW}$&
%$\hat \mu_{AIPW}$&
%$\hat \mu_{REG}$&
%$\hat \mu_{IPW}$&
%$\hat \mu_{AIPW}$
%\\
%\midrule 
%\\
%(500, 1000) &93.6 & 92.0  & 93.0 & 93.0 & 94.2  & 93.2
%\\                    
%       
%(500, 2000)  &93.2  &91.2  & 92.2 & 92.6 & 92.4 &  92.6
%\\         
%       
%(2000, 1000) &94.2&  95.4  & 94.0 & 95.6  &95.8&   95.8
%\\
%        
%(2000, 2000)  &94.2 & 95.0  & 94.4 & 94.6 & 95.0 &  94.6
%\\
%
%               
%\bottomrule
%\end{tabular*}} 
%\end{center}
%\end{table}

\section{An Application to the ESPACOV Survey Data\label{Section4}}

We apply our proposed estimation methods to data collected by  the ESPACOV survey
(Estudio Social sobre la Pandemia de COVID-19) conducted by the Institute for Advanced Social Studies at the Spanish National Research Council \citep[IESA-CSIC]{rinken2020combined}.
The survey was designed to assess the impact of the COVID-19 pandemic in Spain and was conducted from January 18 to 25, 2021, approximately one year into the pandemic. It utilized an online platform and employed a mixed multiphase sampling strategy, which included sending Short Message Service invitations to randomly generated mobile phone numbers (probability-based sample; $\mathcal{S}_B$) and advertising on Facebook, Instagram, and Google Ads (non-probability sample; $\mathcal{S}_A$). Detailed information on the sampling design and data collection can be found in \cite{rinken2020combined} and \cite{rueda2023enhancing}.

Our study focuses primarily on investigating the self-assessment of mood among Spanish residents aged 18 and older during the COVID-19 crisis. The original responses were categorized into five mood levels: very bad, bad, neither bad nor good, good, and very good. For analytical purposes, we convert the responses into a binary variable, assigning a value of 1 to indicate a good mood (including ``good" and ``very good") and 0 otherwise. We consider eight covariates that could potentially influence mood self-assessment and respondents' participation in the non-probability sample; detailed descriptions are provided in Table \ref{dat:variable}.  The age variable was categorized into three groups, with the 18-29 age group serving as the reference category. After excluding missing data, our analysis includes 881 observations from the probability sample and 584 from the non-probability sample.

\medskip
\begin{table}[!htt]
\caption{Descriptions of variables in the ESPACOV survey.}
\centering\label{dat:variable}

\begin{threeparttable}
\setlength{\tabcolsep}{7pt}{
\begin{tabular}{lllll}
\toprule 
Notation&Variable & Level & $\mathcal{S}_A$  &$\mathcal{S}_B$ 
\\
 \midrule
 
$x_1$&Age$^\dag$& 18-29         & $3.3 \%$   & $18.2\%$   \\
       & & 30-44         & $15.8 \%$ & $33.0\%$  \\
       & & 45-64        & $37.4 \%$ & $41.3 \%$    \\
       & & 65 or more & $43.5 \%$ & $7.5 \%$    \\

$x_2$&Education level  & First/second degree    (0)     & $46.8 \%$ & $39.7 \%$  \\
                  &         & Higher education (1)                & $53.2 \%$ & $60.3 \%$  \\

$x_3$& Gender & Male     (0)        & $40.7 \%$ & $48.4 \%$   \\
           &  & Female (1) & $59.3 \%$ & $51.6 \%$  \\

$x_4$&Score of government action$^\ddag$  &  0,1,2,3,4,5 (0)  &61.1\%&66.5\%\\
		                               &	                       & 6,7,8,9,10 (1)    &38.9\%&33.5\%\\

$x_5$&Cost of obeying the policy$^\bot$  &Nothing or a bit (1)         &62.5\%&59.2\%\\
	  &	                    				&Others  (0)			&37.5\%&40.8\%\\

$x_6$&When to get vaccinated 		&Next year or Never (1) &9.9\% &16.7\%\\
	        &						&Others (0)		     &90.1\%&83.3\%\\
	        
$x_7$&Social status self-assessment      & Low/very low  (1)         &31.8\%&30.8\%\\
		&	  & Others  (0) %Middle/high/very high 
		&68.2\%&69.2\%\\

$x_8$&Health self-assessment & Good/Very good (1)  &$76.1\%$        &77.8\%\\
                            &                     & Others   (0)                &$23.9\%$       &22.2\%\\

$y$&Mood self-assessment  &Good/Very good (1)&$43.9\%$  \\
                          &                   &Others   (0)             &$56.1\%$  
\\ 
\bottomrule
\end{tabular}}
\begin{tablenotes}
\item[$^\dag$] Age groups are converted into three dummy variables $x_{11}$, $x_{12}$ and $x_{13}$  
\item[$^\ddag$] Score of the action of the government of Spain to control the pandemic
\item[$^\bot$]
Personal cost of complying with the measure ``limit the number of people in family/friends gatherings'' during the pandemic
\end{tablenotes}
\end{threeparttable}
\end{table}
 
Table \ref{dat:variable} shows noticeable disparities in the distributions of auxiliary variables between the probability and non-probability samples, particularly for demographic variables such as age, gender, and education level. 
%These discrepancies indicate that the two samples were obtained using different sampling methods. 
To address the issue of low response rates typically associated with random digit dialling surveys, the observed data from the probability sample were weighted using iterative raking adjustments for relevant variables.  This weighting procedure has proven to be effective in correcting biases in the ESPACOV survey and previous surveys conducted by IESA-CSIC \citep{rueda2023enhancing}.

The first main objective of our study is to investigate the participation mechanism for the non-probability sample. To achieve this, we utilize health self-assessment as an instrumental variable and model the participation probability (the propensity score) through a logistic regression: 
\[
\pi^A(x, y; \theta) = \Big\{1 + \exp \Big(\alpha + \sum_{j=1}^3 \beta_{1j} x_{1j} + \sum_{j=2}^7 \beta_{j} x_{j} + \gamma y \Big)\Big\}^{-1} \,.
\]
Noting that the response variable $y$ (``experiencing good mood'') is binary, we consider another logistic regression for the outcome regression model:
\[
\pr(y=1 \mid x, R=1) = \frac{\exp \left(\xi_0 + \sum_{j=1}^3 \xi_{1j} x_{1j} + \sum_{j=2}^8 \xi_{j} x_{j} \right)}{1 + \exp \left(\xi_0 + \sum_{j=1}^3 \xi_{1j} x_{1j} + \sum_{j=2}^8 \xi_{j} x_{j} \right)}.
\]

Based on the  two models described above, we compute $\hat\xi$ and $\hat\theta$ along with their corresponding SEs, as shown in Table \ref{dat:est}. 
Our findings indicate that, at the 5\% significance level,  
variables such as age group ($x_{11}$--$x_{13}$) and education level ($x_{2}$) significantly influence participation in the non-probability survey. Additionally, variables ($x_3$--$x_8$) have significant effects on mood. Middle-aged and elderly respondents with higher education levels are more likely to participate in the non-probability survey compared to others. These results align with previous public health studies, which found that self-protection and social motivations can promote participation of middle-aged and older adults \citep{cao2022motivational}, while individuals with higher levels of education are more likely to take part in surveys compared to those with lower educational attainment \citep{spitzer2020biases}. 
Male participants in the non-probability sample who (i) express satisfaction with government actions, (ii) have reservations about fully adhering to policies, (iii) delay vaccination plans, (iv) report good health, and (v) perceive themselves as having high social status, are often associated with good mood. The insignificance of the coefficient $\gamma$ is an unexpected result, which could be attributed to the small sample sizes. A similar observation is noted in Section 3.2, particularly in Table \ref{tab:simBinaryPointEstTheta}, where $\gamma=-0.8$. In that scenario, with $E(n_A)=500$ and $n_B=1000$, the SD of $\hat\gamma$ reaches 0.8, suggesting that   detecting the significance of $\gamma$ may be challenging. However, 
as $E(n_A)$ increases to 2000, the significance of $\gamma$ may become apparent. 
Psychological research indicates that a good mood enhances cooperation, which in turn increases survey participation \citep{carlson1988positive, wolff2022day}. From a psychological perspective, we proceed with our analysis while considering the nonignorable participation mechanism. 

\medskip
 \begin{table}[!htt] 
\caption{Estimated regression coefficients in participation and outcome regression models.}
\centering\label{dat:est}

\begin{threeparttable}
\setlength{\tabcolsep}{12pt}{
\begin{tabular}{lccc | lccc}
\hline
&Estimator & SE & p-value  &   &Estimator & SE & p-value 
\\
 \hline
& \multicolumn{3}{c|}{Participation probability model}&
& \multicolumn{3}{c}{Outcome regression model}\\
$\alpha$      &13.506 &0.365 &0.000&$\xi_0$   &-2.349 &0.558 &0.000 \\
$\beta_{11}$&-0.728 &0.287 &0.011&$\xi_{11}$&0.101 &0.520 &0.847 \\
$\beta_{12}$&-1.498 &0.279 &0.000&$\xi_{12}$&0.384 &0.479 &0.423 \\
$\beta_{13}$&-2.522 &0.316 &0.000&$\xi_{13}$ &0.660 &0.479 &0.168 \\
$\beta_{2}$  &-1.380 &0.177 &0.000&$\xi_2$     &-0.037 &0.192 &0.846 \\
$\beta_{3}$  &-0.131 &0.178 &0.461&$\xi_3$     &-0.478 &0.192 &0.013 \\
$\beta_{4}$  &0.313  &0.189 &0.097&$\xi_4$     &0.556  &0.196 &0.004 \\
$\beta_{5}$  &-0.139 &0.210 &0.508&$\xi_5$     &0.577  &0.202 &0.004 \\
$\beta_{6}$  &0.119  &0.236 &0.615&$\xi_6$	&0.837   &0.325 &0.010 \\
$\beta_{7}$  &-0.257 &0.217 &0.237&$\xi_7$	&-0.461  &0.220 &0.036 \\
$\gamma$   &-0.538 &0.732 &0.463&$\xi_8$	&1.748   &0.260 &0.000 \\  
\bottomrule
\end{tabular}} 
\end{threeparttable}

\end{table}

The second main objective of our study is to estimate the population proportion of Spaniards experiencing good moods. Using the model parameter estimates obtained earlier, we compute the estimated proportion using regression, IPW, and AIPW methods. The resulting estimates are 30.6\%, 30.0\%, and 31.2\%, with corresponding SEs 14.1\%, 14.4\%, and 14.0\%, respectively. 
In contrast, the estimates derived from \cite{chen2020doubly}'s regression, IPW, and AIPW methods are 41.2\%, 40.4\%, and 41.7\%, which exceed our estimates by approximately 32\%. This substantial difference can be attributed to their reliance on the ignorable assumption for the participation mechanism, an assumption that may be questionable. %{\color{red} given that the maximum pseudo-likelihood estimate of the coefficient $\gamma$ in Table \ref{dat:est} significantly deviates from zero.} 
Overall, these estimates consistently fall below the naive estimate 43.9\% observed in the non-probability sample.

\section{Additional Remarks\label{Section5}}

Nonignorable participation mechanism is an important but difficult topic for analysis of non-probability survey samples. 
We developed a pseudo-likelihood estimation method to adjust for selection bias in nonignorable non-probability survey samples under the popular two-sample setup where auxiliary information is available from an existing reference probability sample. Using the maximum pseudo-likelihood estimator of participation probabilities, we constructed the regression-based prediction, IPW, and AIPW estimators for the population mean and studied their asymptotic properties. We also proposed plug-in variance estimators for the three methods. 

The effectiveness of our proposed methods relies on parametric assumptions on models for the participation mechanism and the outcome regression as in \cite{kim2023empirical}. Recent studies, including those on kernel matching \citep{wang2020improving} and Bayesian additive regression trees \citep{rafei2020big, rafei2022robust}, have explored nonparametric assumptions for ignorable participation mechanisms. Extending our methods to using nonparametric models for the nonignorable participation mechanism is a promising direction for future research.

\setcounter{equation}{0}
\def\theequation{A.\arabic{equation}}

\section*{Appendix: Extending the Method of Kim and Morikawa}

We extend the method of \cite{kim2023empirical} to our current two-sample setup. The estimation procedure consists of two  steps. In the first step, a calibration estimator of the regression parameter $\theta$ is obtained by solving the following system of estimating equations: 
\begin{equation} \label{eq:cal-theta}
g(\theta) = 
\sum_{i\in \mS_A} \frac{1}{\pi^A(x_i, y_i; {\theta})} 
\left(
\begin{matrix}
1\\
x_i
\end{matrix}
\right) -
\sum_{i\in \mS_B} d_i
\left(
\begin{matrix}
1\\
x_i
\end{matrix}
\right)
= 0 \,.
\end{equation}
Numerically, the calibration estimator is calculated by minimizing the objective function $\breve\ell(\theta)=g(\theta)^\T g(\theta)$. 

Once the minimizer $\breve\theta=\arg\min\{\breve\ell(\theta)\}$ is obtained, we proceed to the second step by using the approach proposed by  \cite{kim2023empirical} to estimate the population mean through the EL weighting method. The weights are determined by maximizing
$
\sum_{i\in \mS_A} \log(p_i) 
$  
subject to the constraints $\sum_{i\in \mS_A} p_i=1$ and 
\begin{equation}\label{con2}
\sum_{i\in \mS_A} p_i \pi^{A}(x_i, y_i; \breve\theta) = \frac{1}{\sum_{i\in \mS_B} d_i}
\sum_{i\in \mS_B} d_i \pi(x_i; \breve \theta, \hat\xi),
\end{equation}
\begin{equation}\label{con3}
\sum_{i\in \mS_A} p_i x_i = \frac{1}{\sum_{i\in \mS_B} d_i}\sum_{i\in \mS_B} d_i x_i,  
\end{equation}
where $\pi(x;  \theta, \xi) = P(R=1\mid x)$ is defined in \eqref{def.pi.fun}. Condition \eqref{con2} serves as  the bias calibration condition and Condition \eqref{con3} is the benchmarking constraint used in  \cite{kim2023empirical}. Once the maximizer $\breve p_i$ is obtained, the EL estimator of the population mean is computed as $\hat \mu_{EL} = \sum_{i\in \mS_A}\breve p_i y_i$. %By replacing $\breve \theta$ with the maximum pseudo-likelihood estimator $\hat\theta$ in Constraints \eqref{con2} and \eqref{con3}, we defined an alternative EL estimator, denoted by $\hat \mu_{EL\_PL}$ in our simulation studies. 

\section*{Supplementary material}
The supplementary material document
contains the proof of Proposition \ref{prop.conditional}, the regular conditions, and the proof of Theorem \ref{thm-asy}.

\bigskip

\bibliographystyle{abbrvnat} %abbrvnat unsrtnat automatica
\bibliography{mybibfile}

\end{document}